# Transient non-collinear magnetic state for all-optical magnetization switching

*Sergii Parchenko[1, 2, 3], Antoni Frej[4], Hiroki Ueda[5, 6], Robert Carley[3], Laurent Mercadier[3], Natalia Gerasimova[3], Giuseppe Mercurio[3], Justine Schlappa[3], Alexander Yaroslavtsev[3], Naman Agarwal[3, 7], Rafael Gort[3], Andreas Scherz[3], Anatoly Zvezdin[8, 9], Andrzej Stupakiewicz[4], Urs Staub[5].*

1. Laboratory for Mesoscopic Systems, Department of Materials, ETH Zurich, 8093 Zurich, Switzerland
2. Laboratory for Multiscale Materials Experiments, Paul Scherrer Institute, 5232 Villigen PSI, Switzerland
3. European XFEL, Holzkoppel 4, 22869 Schenefeld, Germany
4. Faculty of Physics, University of Bialystok, 1L Ciolkowskiego, 15-245 Bialystok, Poland.
5. Swiss Light Source, Paul Scherrer Institute, 5232 Villigen, Switzerland
6. SwissFEL, Paul Scherrer Institut, 5232 Villigen, Switzerland
7. Department of Physics and Astronomy, Aarhus University, 8000 Aarhus, Denmark
8. Prokhorov General Physics Institute of the Russian Academy of Sciences, Vavilova 38, 119991 Moscow,
9. Russia - New Spintronic Technologies, Bolshoy Bulvar 30, bld. 1, 121205 Moscow, Russia.

Correspondence should be addressed to: sergii.parchenko@xfel.eu (S.P.); and@uwb.edu.pl (A.S.); urs.staub@psi.ch (U. S.).

**Resonant absorption of a photon by bound electrons in a solid can promote an electron to another orbital state or transfer it to a neighboring atomic site. Such a transition in a magnetically ordered material could affect the magnetic order. While this process is an obvious road map for optical control of magnetization, experimental demonstration of such a process remains challenging. Exciting a significant fraction of magnetic ions requires a very intense incoming light beam, as orbital resonances are often weak compared to above-band-gap excitations. In the latter case, a sizeable reduction of the magnetization occurs as the absorbed energy increases the spin temperature, masking the non-thermal optical effects. Here, using ultrafast x-ray spectroscopy, we were able to resolve changes in the magnetization state induced by resonant absorption of infrared photons in Co-doped yttrium iron garnet, with negligible thermal effects. We found that the optical excitation of the Co**

**ions affects the two distinct magnetic Fe sublattices differently, resulting in a transient non-collinear magnetic state. The present results indicate that the all-optical magnetization switching most likely occurs due to the creation of a transient, non-collinear magnetic state followed by coherent spin rotations of the Fe moments.**

Ultrashort optical pulses have proven to serve as an efficient stimulus to excite magnetization dynamics [1]. Among the most prominent examples are the demonstration of ultrafast quenching of the magnetization [2, 3], single [4, 5] and multiple [6, 7] pulse all-optical magnetization switching, optically induced magnetic phase transitions [8, 9, 10], and optical excitation of spin-waves [11, 12, 13]. In these cases, the investigations aimed to understand the ultrafast dynamics initiated by significant energy deposition to the system, also causing a reduction of the magnetization. On the other hand, ultrafast magnetization manipulation methods that do not cause magnetization suppression are much less explored. While there are several successful demonstrations of non-thermal magnetization manipulations reported covering different frequency ranges of photo excitations [14, 15, 16, 17, 18, 19], studies showing a permanent magnetization change, such as magnetization switching, are rare [20, 21]. Among the materials demonstrating non-thermal magnetization excitation, Co-doped yttrium iron garnets (YIG:Co) attracted special interest due to the possibility of very efficient magnetic state manipulation using site-selective optical excitations [22]. The growing interest is motivated by the demonstration of deterministic switching of the magnetization direction by a single femtosecond near-infrared (NIR) laser pulse [23]. Additionally, the induced direction of the magnetization can be controlled by the linear polarization of the optical driving pulse. While the excitation process is attributed to the change of the magnetic anisotropy caused by the photomagnetic effect [24], the microscopic mechanism underlying the magnetization dynamics in different magnetic sublattices remains unknown.

Typically, materials that exhibit strong photomagnetic effects contain ions that are not significantly contributing to the size of the macroscopic magnetic moment but may affect the magnetic properties due to their spin-orbital coupling [15]. After the photoexcitation, changes in the electronic and magnetic state of these ions can induce a change of the direction of the easy magnetization axis, triggering the change of the magnetization orientation [25]. One of the key criteria to obtain a strong photomagnetic effect is the presence of an optically active electronic transition of the impurity ions in the spectral range where the material is otherwise optically transparent. YIG:Co nicely fulfills this criterion. The system used in our study has the chemical formula $Y_2CaFe_{3.9}Co_{0.1}GeO_{12}$ – a cubic ferrimagnet without a magnetic compensation point and a Curie temperature of $T_C$=445 K. The sample is an 8.5-μm thick film on top of gadolinium gallium garnet substrate with [001] out-of-plain direction. Two antiparallel magnetic sublattices are formed by Fe ions in tetrahedral and octahedral oxygen coordination. Tetrahedral and octahedral $Fe^{3+}$ ions are unequally substituted by $Ge^{4+}$ ions that significantly decrease the net magnetization ($4\pi M_S$=56 G). Preferred magnetization orientations are along the <111> directions. Trivalent Fe ions with a half-filled $d$ shell have a negligible orbital magnetic moment. In contrast, $Co^{2+}$ and $Co^{3+}$ ions that substitute Fe in both crystallographic sites [26, 27, 28], have significant orbital magnetic moments, which strongly affect the magnetic configuration of the material. The pure yttrium iron garnet is optically transparent for wavelengths above 900 nm as all electronic transitions of the Fe ions lie in the visible range [29]. The substituted Co ions add several narrow electronic transitions in the NIR range [30, 31]. The proposed trigger for the photoinduced magnetization excitation is based on the resonant excitation of a localized $d$-$d$ transition in Co-ions [22]. Optical excitation alters the orbital state of Co, which in turn affects the soft Fe moments, resulting in a rotation of the magnetization vector to align with another equivalent <111> direction.

Optical methods are widely used to investigate macroscopic ultrafast photo-magnetic dynamics [1]. However, revealing the microscopic picture of the magnetization excitation after an electronic transition in a coupled system requires information about the dynamics of the individual magnetic sublattices. To gain that information, we performed time-resolved NIR pump and soft x-ray probe experiments. We employ soft x-ray magnetic circular dichroism (XMCD) in reflectivity mode [32] to determine the magnetization dynamics of the Fe moments occupying octahedral and tetrahedral sites separately.

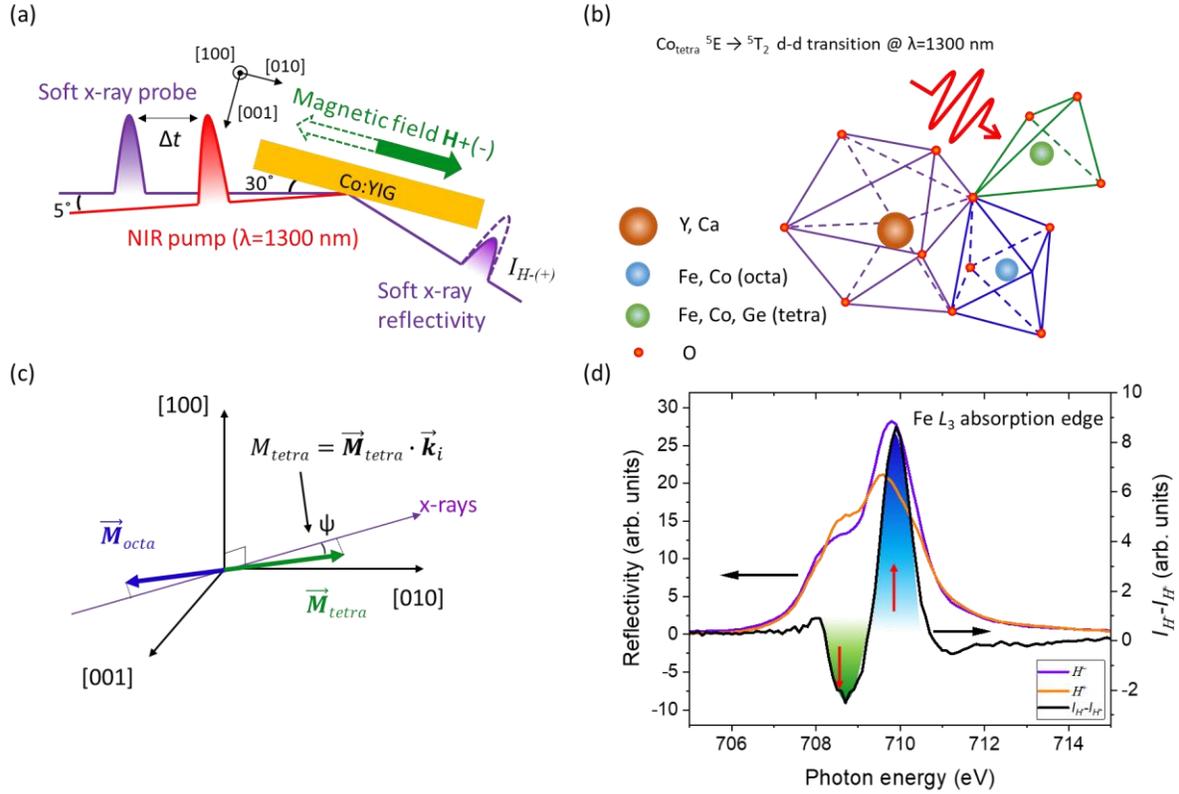

*Figure 1. **Geometry and probing method of the time-resolved XMCD experiment.** (a) Experimental geometry of the time-resolved XMCD experiments. The x-ray incidence angle is 30° and the angle between the pump and probe beams is 5° (b) Sketch representing the crystallographic occupation of different ions in YIG:Co films. Here, blue and green color coding is used to show octahedral and tetrahedral crystallographic sites respectively. The same color coding is used throughout the manuscript for the data. (c) Visualization of the probing components. Magnetic moments are close to the [010] direction in the initial state. (d) Static x-ray reflectivity around the Fe $L_3$-absorption edge for opposite orientations of the in-plane magnetic field and its difference. Filled areas with different colors show the energy range where the dominant contribution is from different magnetic sublattices: green - $M_{tetra}$ and blue – $M_{octa}$. Red arrows point to energies used in time-resolved experiments.*

The experimental geometry is shown in Fig. 1. Further experimental details are described in the Methods and Supplementary material sections. The reflectivity XMCD method probes a projection of the magnetic moment onto the incoming x-rays, which is, to the first order, the scalar product of the magnetization and the $\mathbf{k}_i$ vector of the x-ray probe for small incident angles. Even though this approximation is not well satisfied in our experiment, the signal remains sensitive to the magnetization direction. The Fe magnetic moments in tetrahedral and octahedral sites can be separated in the soft x-ray range due to the different crystal field potentials of their sites, which results in distinct spectral features in the Fe $L_3$-edge x-ray absorption spectrum [33]. The Fe $L_3$-edge reflectivity spectra for opposite directions of the external magnetic field are shown in Fig. 1d. The difference reflectivity signal from oppositely aligned $M_{tetra}$ and $M_{octa}$ magnetic sublattices have a spectral maximum and minimum at $E_{tetra}$=708.5 eV and $E_{octa}$=709.8 eV, respectively. Such energy separation is sufficient to determine the individual magnetization behavior with the available energy resolution [34].

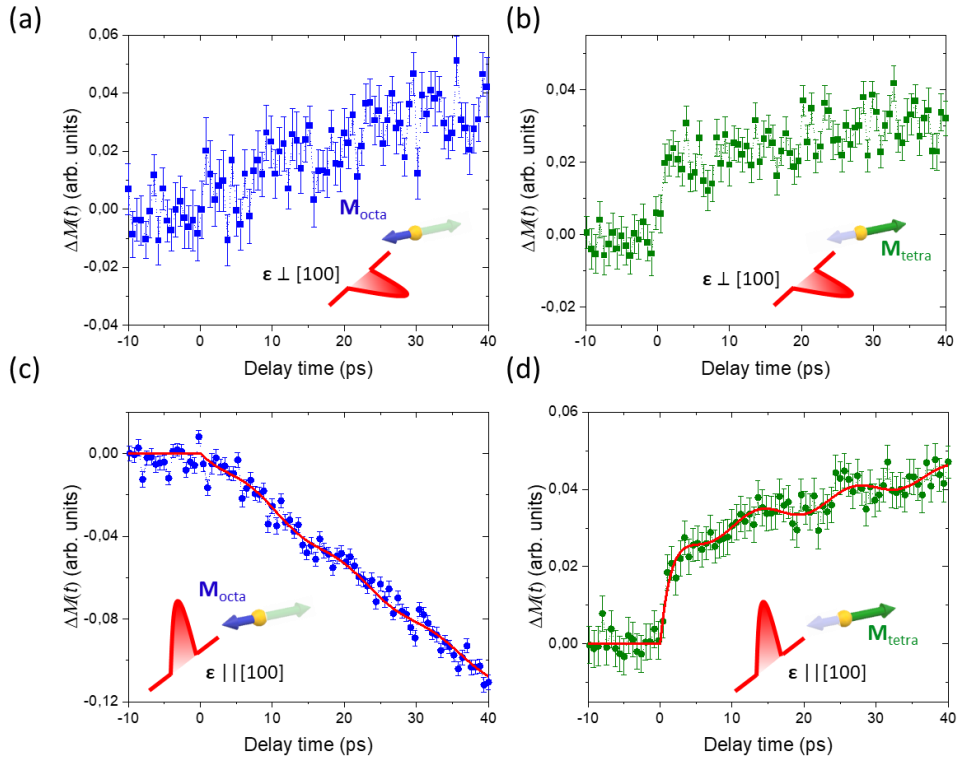

*Figure 2. **Photo-induced time-resolved magnetization dynamics in YIG:Co.** Time-resolved changes in magnetization projection for octahedral ((a) and (c)) and tetrahedral ((b) and (d)) Fe sublattices, excited with pump pulses with different orientations of linear polarization ε. Insets show a sketch for the pump polarization configuration and probed magnetization component at each panel. Red solid lines at panels (c) and (d) are fits to Eq. 3. For both pump polarizations, a fast component in the magnetization dynamics of the tetrahedral magnetic sublattice is observed, which is absent for the octahedral sublattice.*

The laser-induced dynamics of the magnetic signal are due to the change of the magnetization direction for the respective sublattice as the thermal effects are negligible at 1300 nm excitation wavelength [23]. Thus, the time-resolved magnetization dynamics is better described in terms of transient change of magnetization vector orientation with respect to the x-rays (see Methods and Supplementary materials for more details). We clarify here that after the optical excitation we do not achieve the single-shot laser-induced magnetization switching because of the external magnetic field. However, the initial optically induced photomagnetic changes are expected to be independent of the external magnetic field. Time-resolved magnetization dynamics measured with the optical probe method in the same configuration as during the experiment with the x-ray probe, displaying a longer delay range, are shown in SFig. 3. The magnetization dynamics for the individual magnetic sublattices measured with the x-ray probe differ substantially from each other. The $\mathbf{M}_{octa}$ moments follow a regular magnetization precession behavior consistent with the ferromagnetic resonance (FMR) precession mode that could be described by the Landau-Lifshitz-Gilbert formalism, similar to what is observed with optical probing methods [24]. In contrast, the dynamics of $\mathbf{M}_{tetra}$ show a prompt change (within a picosecond) after the optical excitation, which we refer to as the picosecond magnetization (PM) component. The fast component of the magnetization dynamics is present in the data for both linear pump pulse polarizations (see Fig. 2b and d). The disparate dynamics for the two initially antiparallel magnetic moments mean that the optical excitation creates a transient non-collinear magnetic state. The characteristic time constant of the PM component is $\tau_{PM}=1.6\pm0.7$ ps. It is worth noticing that the fast PM component in $M_{tetra}$ is responsible for about half of the $\Delta M$ signal change when comparing the equilibrium state with the delay time $\Delta t=40$ ps, which corresponds approximately to the magnetization switching time in YIG:Co [23] For the two nearly compensated antiferromagnetically aligned moments in YIG:Co, even a tiny change of magnetization vector pointing of one sublattice significantly affects the total magnetization $\mathbf{M} = \mathbf{M}_{tetra} + \mathbf{M}_{octa}$. After the excitation, the two magnetic moments are no longer antiparallel and the vector sum is increased. Therefore, the optical generation of a non-collinear state result in a significant change of the magnetization orientation and effectively means a transient increase of the magnetization $M$. The non-collinear transient state observed after laser excitation seems crucial for all-optical switching in YIG:Co.

Concomitant with the prompt response, we observe weak oscillations that can be attributed to the quasi-antiferromagnetic resonance mode (q-AFMR) [35]. While the q-AFMR dynamics at $M_{octa}$ are hardly visible, more clear oscillatory behavior is seen in $M_{tetra}$. From a fit (see Methods), we obtain frequencies of 77 ± 6 GHz and 84 ± 9 GHz for the dynamics of tetrahedral and octahedral magnetic sublattices, respectively. The obtained values are, within the uncertainty range, in good agreement with the theoretical prediction (see Supplementary materials).

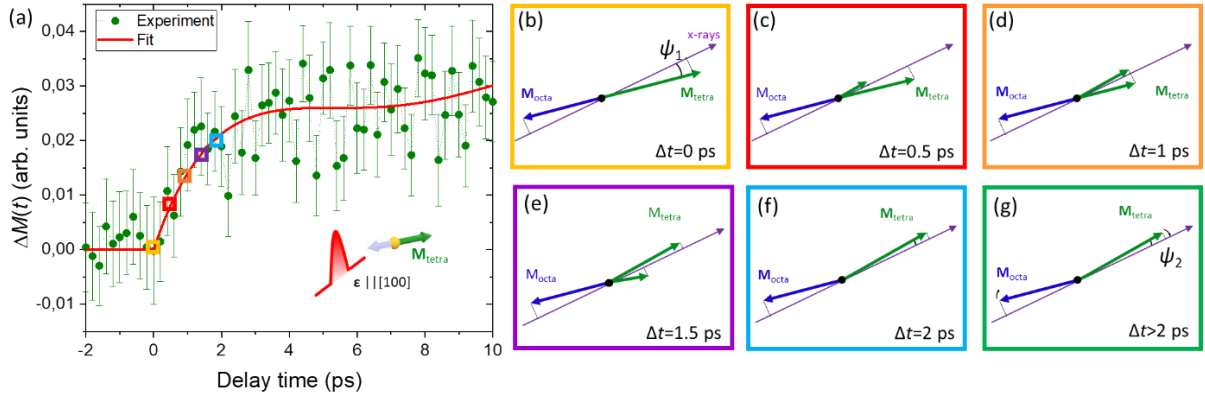

*Figure 3. **Ultrafast dynamics of the non-collinear magnetization state.** (a) Time-resolved changes of magnetization projection at early delay times of tetrahedral sublattice ΔM onto the x-rays when excited with optical pump polarization parallel to [100] direction. The red line is a fit (see methods section). Color squares correspond to times where panels (b) – (f) show sketches of the evolution of magnetization orientation with respect to the x-ray probe. Panel (b) shows the magnetic state before the excitation. During the first 2 ps after the excitation, $M_{tetra}$ along the equilibrium state shrinks and simultaneously grows along the new preferred orientation ((c)-(f)). Later dynamics are determined by q-AFRM and FMR modes. Here, both magnetic moments rotate clockwise resulting in a decreasing projection component for $M_{octa}$ and an increasing for $M_{tetra}$ as the laser excitation drives $\mathbf{M}_{tetra}$ to the other side of $\mathbf{k}_i$ as shown in panel (g).*

The disparate dynamics of the two magnetic sublattices is striking as one would expect both magnetic sublattices to follow the pump-induced change of orbital occupancy of Co-ions similarly as a macrospin dynamics. However, the signs of the Δ*M* dependencies are inconsistent with simple rotational magnetization dynamics, even though we can't determine the exact magnetization orientation after the photoexcitation. In our data, positive Δ*M* signal indicates that the magnetic moment of the sublattice is moving towards the incoming x-ray direction $\mathbf{k}_i$, whereas a negative signal implies that the magnetization is moving in the opposite direction. When excited with a laser pulse polarized perpendicular to [100] crystallographic direction both $\Delta M_{tetra}$ and $\Delta M_{octa}$, increase. That indicates that the magnetization of both magnetic sublattices turns towards $\mathbf{k}_i$. However, when excited with a pump pulse polarized along the [100] direction $\Delta M_{tetra}$ increases but $\Delta M_{octa}$ decreases with time. These unusual dynamics continue even after the prompt response of $\Delta M_{tetra}$ to the optical excitation, suggesting that non-collinearity increases with time. Such a process is very unlikely as magnetic moments have to overcome the exchange interaction, which forces moments to be in a collinear antiferromagnetic state. To comply with the acting of the exchange interaction, the instant change of $\Delta M_{tetra}$ after about 2 ps must result in a $\mathbf{M}_{tetra}$ flipping to the opposite side of $\mathbf{k}_i$ of the x-rays (see Fig. 3). We consider only the relative change of the magnetization orientation of each sublattice with respect to the x-ray probe direction. We state that the magnetic moment of $M_{tetra}$ after the excitation shrunk along the initial equilibrium direction and grew along the new laser-defined orientation as shown in Fig. 3. The sum of these two individual component amplitudes remains constant and equal to $M_{tetra}$ at the equilibrium state, but the spatial distribution of moments is inhomogeneous as discussed below. Optical excitation with light polarization along [100] brings the $\mathbf{M}_{tetra}$ to the configuration when $\psi_2<\psi_1$ (see Fig. 3), explaining the transient increase of the $\mathbf{M}_{tetra}$ magnetization vector projection onto the x-rays. During the recovery dynamics angle $\psi$ decreases and $\mathbf{M}_{tetra}$ approaches $\mathbf{k}_i$ causing an increase of the projection component of $\mathbf{M}_{tetra}$, onto $\mathbf{k}_i$ as visualized in Fig. 3. At the same time, the exchange coupling will force $\mathbf{M}_{octa}$ to follow $\mathbf{M}_{tetra}$ resulting in the decrease of its projection component. By this, both moments approach a collinear state at later times.

To comply with the strong force of the exchange interaction with a monotonic increase of the observed non-collinear state during the first 2 ps after the pump pulse excitation, an inhomogeneous time-dependent distribution of $\mathbf{M}_{tetra}$ magnetic moments is required. The trigger of magnetization dynamics of Fe is the photoexcitation of Co ions. The number of Co ions with respect to the number of Fe ions in YIG:Co film is ~ 1:40. Co-ions occupy tetrahedral and octahedral sites equally [26,27], but the optical excitation with 1300 nm light affects mainly tetrahedral Co sites [22]. With slightly less than 0.5 of Co ions per unit cell, the average distance between two Co ions is about twice the lattice constant. When assuming that all tetrahedral Co ions in the probed area are excited, the two nearest excited Co ions will be separated by at least several Fe ions. The optical excitation of a Co ion changes its orbital configuration, which in turn affects first the nearest Fe moments before the ones further away. Such an excitation process creates a time-dependent distribution of the "polaronic type" [40] spin texture of Fe moments as sketched in figure 4. The polarons then grow and a homogenous state, characterized by regular spin dynamics, is achieved. Magnetization dynamics at later times are then determined by magnetization precession around the "easy" axis in the equilibrium state due to the action of the external magnetic field and defined by the ferromagnetic resonance mode.

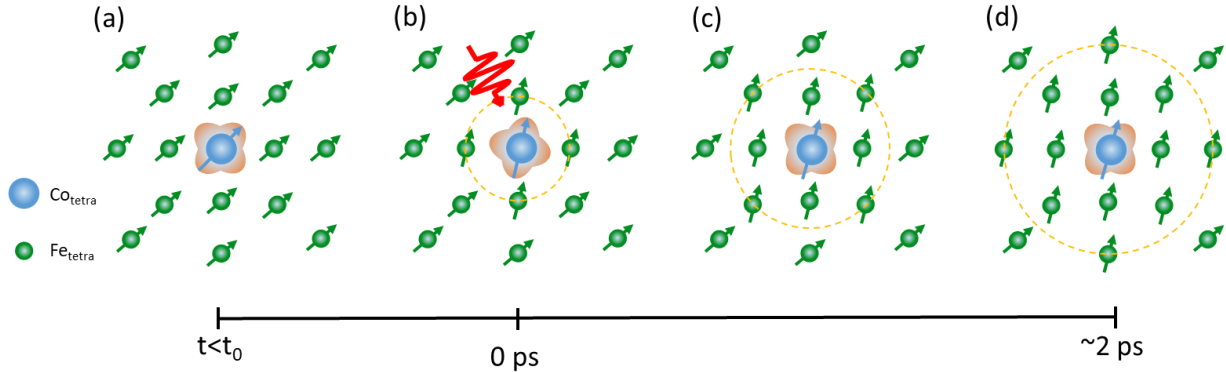

*Figure 4. **Schematic visualization of microscopic magnetization dynamics.** Here, only magnetic ions in tetrahedral sites are shown. Prior to excitation, all $\mathbf{M}_{tetra}$ moments are along the equilibrium orientation direction (a). Optical excitation changes the orbital state of the Co ion, which results in a different single-ion anisotropy easy-magnetization axis. This induces the reorientation of spins of the nearest Fe moments (b). The reorientation cascade, shown as a dashed orange circle, propagates away from the initially excited Co ion affecting Fe ions located further away as shown on panels (c) and (d).*

We note that the excitation pump fluence in our experiments is above the threshold for photo-magnetic switching [23]. Therefore, we assume that photo-magnetic switching occurs when polaronic spin cascades from neighboring Co ions merge. This implies two conditions that must be met to achieve stable switching: firstly, the excitation fluence must be high enough to ensure that a sufficient number of Co ions are excited, launching cascades that eventually merge (this is consistent with the relatively high threshold laser fluence for switching); and secondly, the area on the sample where the first condition is fulfilled must be large enough to allow a temporarily switched domain to stabilize.

By clarifying the mechanism of photomagnetic magnetization switching, we were able to demonstrate that resonant absorption of a photon by a small number of Co-ions can drastically change the magnetic state of many surrounding Fe-ions. The observed spin-polaronic cascade might be a tool to transiently control the spin texture on the atomic level of many other materials with weak anisotropies on ultrafast timescales creating an alternative route for ultrafast magnetization control.

**Methods**

**Material.** An 8.5 µm thick single crystal film of $Y_2CaFe_{3.9}Co_{0.1}GeO_{12}$ was grown by liquid phase epitaxy on Gadolinium Gallium Garnet (GGG) with [001] out-of-plain direction.

**Static x-ray characterization**

Static reflectivity XMCD characterization around the Fe $L_3$-absorption edge has been performed at the RESOXS endstation [36] at the SIM beamline [37] of the Swiss Light Source. Experiments have been performed at room temperature. Circularly polarized x-rays reached the sample at 5º from the sample plane. Energy-dependent reflectivity signals were measured in specular reflection geometry with Si photodiode at 10º with respect to the sample surface. The magnetic field generated by the permanent magnet was applied along the [010] direction (which is similar to the time-resolved experiments) to ensure a monodomain magnetic state.

**Time-resolved experimental setup.**

The time-resolved soft x-ray experiment was performed at the SCS instrument of the SASE3 beamline of the European X-ray Free-Electron Laser (FEL) [38]. The experimental geometry is sketched in Fig. 1a. The experiments were performed at room temperature. Optical pump excitation was done with linearly polarized ultrashort laser pulses with $\lambda$=1300 nm wavelength and pulse duration of 100 fs. The sample absorbs about 12% at this wavelength [23]. The optical pump matches a $^5E \rightarrow ^5T_2$ *d-d* transition in the Co ions occupying the tetrahedral site [31]. The pump beam was focused to 350×200 µm spot on the sample at a 35° incident angle with a fluence of 60 mJ/cm$^2$. The linear polarization of the pump pulses was either parallel or perpendicular to [100] direction (see Fig. 1). The x-ray beam was monochromatized using a grating monochromator of the SASE3 beamline [34]. The circularly polarized x-ray pulses were obtained by transmitting FEL-generated x-rays with linear polarization through a perpendicularly magnetized metallic Fe thin film polarizer. Due to the XMCD effect in the Fe film, the transmitted x-ray beam is nearly circularly polarized. [39] An incidence angle of the x-ray probe was set to 30° with respect to the sample surface with a spot size of 100×50 µm on the sample. The pulse duration of the x-ray probe was about 35 fs. Pump pulses were arriving with a 56.5 kHz repetition rate and the probe pulses with 113 kHz. This configuration was chosen to increase the signal-to-noise ratio by taking the ratio of pumped and unpumped signals. Incoming x-ray intensity $I_0$ was measured using an x-ray gas-monitor detector. The reflectivity signal $I_1$ was collected with an x-ray Si photodiode. Magnetization-induced reflectivity differences were accessed by an alternating external magnetic field of $H=\pm 250$ Oe applied in the sample plane along the [010] crystallographic direction. This method gives equivalent information about the magnetization state as reversing the chirality of circularly polarized x-rays typically used in XMCD studies (see SFig. 2). Note that due to the early stage of the instrument operation reversing the x-ray chirality was not yet implemented.

As the magnetic signal interferes in the reflection geometry of the experiment with a dominating charge signal, magnetization contrast makes the dichroic signal mainly proportional to the magnetic scattering factor. This is a common way to obtain a signal proportional to the magnetization of the ion when more direct absorption measurements are not feasible. It however restricts the possibility to apply XMCD sum rules as both real and imaginary magnetic scattering factors contribute. As the change in the magnitude of the magnetization vector is negligible the main contribution to the time-resolved magnetic signal is from the change of the orientation of magnetic moments. For sufficiently small incident angles, the XMCD signal is proportional to a magnetization projection onto the incoming resonant x-rays:

$$M \parallel \mathbf{k}_i = cos(\psi)\mathbf{M} \sim I_{H-} - I_{H+} \qquad [1]$$

where $I_{H-(+)}$ is the intensity of the reflected x-ray beam at a given energy in the vicinity of the Fe $L_3$ edge for the opposite direction ($H-/+$) of the external magnetic field, $\psi$ is the angle between the magnetization **M** and the x-ray probe beam. As the incident angle is not small enough, this simple approximation is not fully accurate, and additional moment projections contribute to the signal. These additional contributions prohibit a determination of a quantified moment change, both in direction and size, however, it will not affect the sign of the signal, which is most important in this study. Note that for the static experiments with a small incidence angle, Eq. 1 is a reasonable approximation. In the x-ray energy dependence of the dichroic signal, we can assign the extrema to the magnetization of the tetrahedral and octahedral sites occurring at x-ray energies of $E_{tetra}$=708.5 eV and $E_{octa}$=709.8 eV, respectively, which are the energies chosen for the time-resolved experiment.

The presented dynamics of the magnetization are relative to the equilibrium magnetization orientation. Within the simple approximation, $\Delta M$ corresponds to:

$$\Delta M(t) \sim \frac{I_{H-}^{pump}(t)}{I_{H-}^{unpump}(t)} - \frac{I_{H+}^{pump}(t)}{I_{H+}^{unpump}(t)} \qquad [2]$$

where $I^{pump(unpump)}{}_{H-(+)}$ is the x-ray reflectivity on pumped or unpumped events for the *H*- or *H*+ direction of the external magnetic field.

**Fitting procedure**

To extract the relevant quantities of the photomagnetic effect from the derived experimental data the fit function:

$$\Delta M(t) = A_{PM}\exp\left(-\frac{t}{\tau_{PM}}\right) + A_{q-AFRM}\sin(2\pi f_{q-AFRM}t - \varphi)\exp\left(-\frac{t}{\tau_{q-AFRM}}\right) +$$
$$+ A_{FMR}\sin(2\pi f_{FMR}t - \varphi_{FMR}) \qquad [3]$$

was used. Here PM, q-AFRM and FMR describe picosecond magnetization dynamics, quasi-antiferromagnetic resonance and ferromagnetic resonance modes, respectively. The frequency and phase for the FMR component were fixed to those obtained from time-resolved magneto-optical experiments performed with the same conditions (see Supplementary materials). The obtained values are:

Table 1. Fit values to the dependencies shown in Figures 2 c and d.

| Magnetic sublattice | $A_{PM}$, (arb. units) | $A_{q\text{-}AFRM}$, (arb. units) | $\tau_{PM}$, (ps) | $\tau_{q\text{-}AFRM}$, (ps) | $f_{q\text{-}AFMR}$, (GHz) |
|---|---|---|---|---|---|
| $M_{tetra}$ | 0.029±0.004 | 0.0034±0.0013 | 1.6±0.7 | 55±45 | 77±6 |
| $M_{octa}$ | - | 0.0029±0.0016 | - | 81±32 | 84±9 |

**Acknowledgments:** This work has been funded by the Foundation for Polish Science (Grant No. POIR.04.04.00-00-413C/17). H.U. was supported by the National Centers of Competence in Research in Molecular Ultrafast Science and Technology (NCCR MUST-No. 51NF40-183615) from the Swiss National Science Foundation and from the European Union's Horizon 2020 research and innovation program under the Marie Skłodowska-Curie Grant Agreement No. 801459 – FP-RESOMUS. A.K.Z. acknowledges the Russian Science Foundation (Grant No. 22-12-38700367)

**Author contributions:**

S.P., A.S. and U.S. conceived the project, S.P., A.F., H.U., R.C., L.M., N.G., G.M., J.S., A.Y., N.A., R.G., A.Sch., A.S., U.S. conducted free-electron laser experiment, A.Z. made theoretical calculations. S.P. made the data analysis. S.P., A.S. and U.S. co-wrote the manuscript with important contributions from all co-authors.

**References**

1. Kirilyuk, A., Kimel, A. V. & Rasing, T. Ultrafast optical manipulation of magnetic order. *Rev. of Mod. Phys.* **82**, 2731–2784 (2010).
2. Beaurepaire, E., Merle, J.-C., Daunois, A. & Bigot, J.-Y. Ultrafast Spin Dynamics in Ferromagnetic Nickel. *Phys. Rev. Lett.* **76**, 4250–4253 (1996).
3. Koopmans, B. *et al.* Explaining the paradoxical diversity of ultrafast laser-induced demagnetization. *Nat. Mater.* **9**, 259–265 (2010).
4. Stanciu, C. D. *et al.* All-Optical Magnetic Recording with Circularly Polarized Light. *Phys. Rev. Lett.* **99**, (2007).
5. Ostler, T. A. *et al.* Ultrafast heating as a sufficient stimulus for magnetization reversal in a ferrimagnet. *Nat. Commun.* **3**, 666 (2012).

6. Lambert, C.-H. *et al.* All-optical control of ferromagnetic thin films and nanostructures. *Science* **345**, 1337–1340 (2014).

7. Mangin, S. *et al.* Engineered materials for all-optical helicity-dependent magnetic switching. *Nat. Mater.* **13**, 286–292 (2014).

8. Awari, N. *et al.* Monitoring laser-induced magnetization in FeRh by transient terahertz emission spectroscopy. *Appl. Phys. Lett.* **117**, 122407 (2020).

9. Li, G. *et al.* Ultrafast kinetics of the antiferromagnetic-ferromagnetic phase transition in FeRh. *Nat Commun* **13**, 2998 (2022).

10. Afanasiev, D. *et al.* Control of the Ultrafast Photoinduced Magnetization across the Morin Transition in $DyFeO_3$. *Phys. Rev. Lett.* **116**, 097401 (2016).

11. Kruglyak, V. V. *et al.* Imaging Collective Magnonic Modes in 2D Arrays of Magnetic Nanoelements. *Phys. Rev. Lett.* **104**, (2010).

12. Satoh, T. *et al.* Directional control of spin-wave emission by spatially shaped light. *Nat. Photon* **6**, 662–666 (2012).

13. Deb, M., Popova, E., Jaffrès, H.-Y., Keller, N. & Bargheer, M. Controlling High-Frequency Spin-Wave Dynamics Using Double-Pulse Laser Excitation. *Phys. Rev. Applied* **18**, 044001 (2022).

14. Kimel, A. V. *et al.* Ultrafast non-thermal control of magnetization by instantaneous photomagnetic pulses. *Nature* **435**, 655–657 (2005).

15. Hansteen, F., Kimel, A., Kirilyuk, A. & Rasing, T. Nonthermal ultrafast optical control of the magnetization in garnet films. *Phys. Rev. B* **73**, (2006).

16. Parchenko, S., Stupakiewicz, A., Yoshimine, I., Satoh, T. & Maziewski, A. Wide frequencies range of spin excitations in a rare-earth Bi-doped iron garnet with a giant Faraday rotation. *Appl. Phys. Lett.* **103**, 172402 (2013).

17. Vicario, C. *et al.* Off-resonant magnetization dynamics phase-locked to an intense phase-stable terahertz transient. *Nat. Photon.* **7**, 720–723 (2013).

18. Fitzky, G., Nakajima, M., Koike, Y., Leitenstorfer, A. & Kurihara, T. Ultrafast Control of Magnetic Anisotropy by Resonant Excitation of 4 f Electrons and Phonons in $Sm_{0.7}Er_{0.3}FeO_3$. *Phys. Rev. Lett.* **127**, 107401 (2021).

19. Baierl, S. *et al.* Nonlinear spin control by terahertz-driven anisotropy fields. *Nat. Photon.* **10**, 715–718 (2016).

20. Zhang, P. *et al.* All-optical switching of magnetization in atomically thin $CrI_3$. *Nat. Mater.* **21**, 1373–1378 (2022).

21. Schlauderer, S. *et al.* Temporal and spectral fingerprints of ultrafast all-coherent spin switching. *Nature* **569**, 383–387 (2019).

22. Stupakiewicz, A. *et al.* Selection rules for all-optical magnetic recording in iron garnet. *Nat Commun* **10**, 612 (2019).

23. Stupakiewicz, A., Szerenos, K., Afanasiev, D., Kirilyuk, A. & Kimel, A. V. Ultrafast nonthermal photo-magnetic recording in a transparent medium. *Nature* **542**, 71–74 (2017).

24. Atoneche, F. *et al.* Large ultrafast photoinduced magnetic anisotropy in a cobalt-substituted yttrium iron garnet. *Phys. Rev. B* **81**, 214440 (2010).

25. Kimel, A. V., Kalashnikova, A. M., Pogrebna, A. & Zvezdin, A. K. Fundamentals and perspectives of ultrafast photoferroic recording. *Phys. Rep.* **852**, 1–46 (2020).

26. Maziewski, A. Unexpected magnetization processes in YIG + Co films. *J. Magn. and Magn. Mater.* **88**, 325–342 (1990).

27. Chizhik, A. B., Davidenko, I. I., Maziewski, A. & Stupakiewicz, A. High-temperature photomagnetism in Co-doped yttrium iron garnet films. *Phys. Rev. B* **57**, 14366–14369 (1998).

28. Stupakiewicz, A., Maziewski, A., Davidenko, I. & Zablotskii, V. Light-induced magnetic anisotropy in Co-doped garnet films. *Phys. Rev. B* **64**, 064405 (2001).

29. Landolt-Börnstein: Numerical Data and Functional Relationships in Science and Technology New Series, Group III, 27/e (Springer, 1991).

30. Wood, D. L. & Remeika, J. P. Optical absorption of tetrahedral Co3+ and Co2+ in garnets. J. Chem. Phys. 46, 3595–3602 (1967).

31. Šimša, Z. Optical and magnetooptical properties of Co-doped YIG films. Czech. J. Phys. B 34, 78–87 (1984).

32. Parchenko, S. *et al.* Ultrafast probe of magnetization dynamics in multiferroic $CoCr_2O_4$ and $Co_{0.975}Ge_{0.025}Cr_2O_4$. *Phys. Rev. B* **105**, 064432(2022).

33. Brice-Profeta, S. *et al.* Magnetic order in - nanoparticles: a XMCD study. *J. Magn. and Magn. Mater.* **288**, 354–365 (2005).

34. Gerasimova, N. *et al.* The soft X-ray monochromator at the SASE3 beamline of the European XFEL: from design to operation. *J. Synchr. Rad.* **29**, 1299–1308 (2022).

35. Davydova, M. D., Zvezdin, K. A., Kimel, A. V. & Zvezdin, A. K. Ultrafast spin dynamics in ferrimagnets with compensation point. *J. Phys.: Condens. Matter* **32**, 01LT01 (2020).
36. Staub, U. *et al.* Polarization analysis in soft X-ray diffraction to study magnetic and orbital ordering. *J Synch. Rad* **15**, 469–476 (2008).
37. U. Flechsig, F. Nolting, A. Fraile Rodríguez, J. Krempaský, C. Quitmann, T. Schmidt, S. Spielmann, and D. Zimoch, *AIP Conf. Proc.* **1234**, 319 (2010)
38. Tschentscher, T. *et al.* Photon Beam Transport and Scientific Instruments at the European XFEL. *Appl. Sci.* **7**, 592 (2017).
39. Büttner, F. *et al.* Observation of fluctuation-mediated picosecond nucleation of a topological phase. *Nat. Mater.* **20**, 30–37 (2021).
40. Franchini, C., Reticcioli, M., Setvin, M. & Diebold, U. Polarons in materials. *Nat Rev Mater* **6**, 560–586 (2021).

# Supplementary materials

1) Magnetic Hysterisis Loop

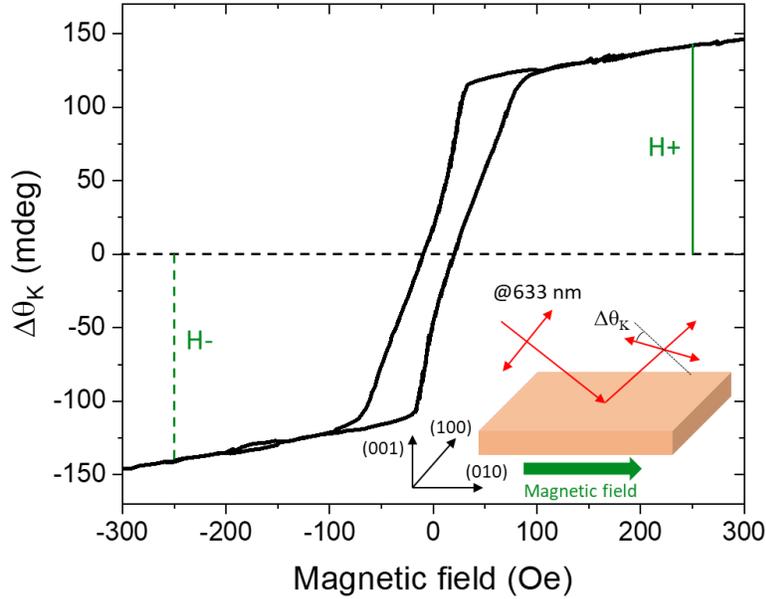

*SFigure 1. Magnetic hysteresis loop measured by longitudinal magneto-optical Kerr effect. The value of the opposite external magnetic field during the time-resolved experiments is indicated with H-(+).*

SFigure 1 shows a magnetic hysteresis loop, recorded using the magneto-optical Kerr effect. The magnetic field was applied in the sample plane parallel to the [010] direction, which is the same configuration as during the time-resolved x-ray experiments. Linearly polarized light from a GaAs laser with wavelength 633 nm impinged the sample at a 45° incidence angle. In this geometry, the in-plane magnetic moment is probed. At an approximately $H$=100 Oe field the sample is already in the monodomain state. However, with further increasing field, the Kerr rotation signal $\Delta\theta_K$ signal increases too, indicating that the magnetization is not exactly in the plane but has some additional out-of-plane contributions as described in Fig. 1c in the main text.

2) Static x-ray reflectivity

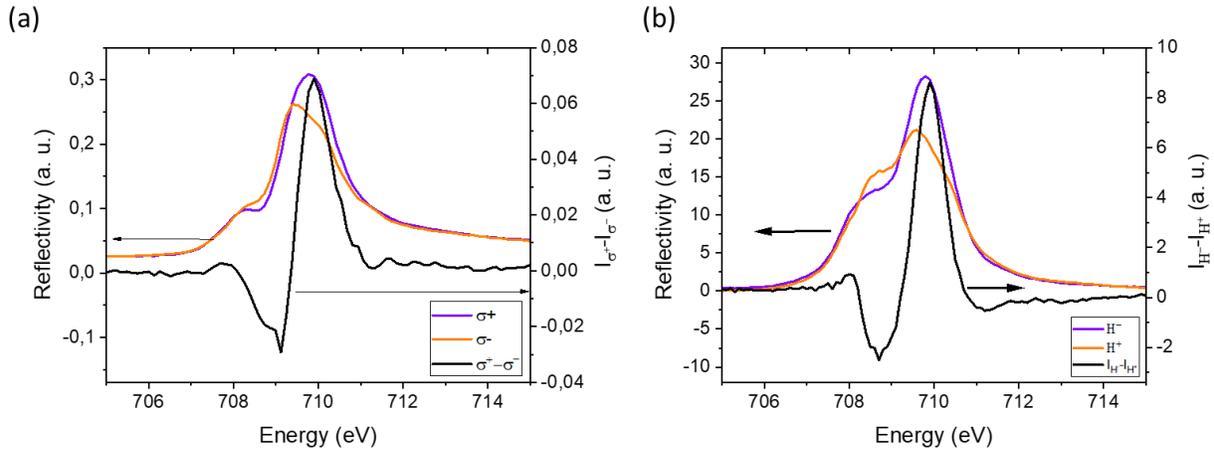

*SFigure 2. Comparison of magnetization-induced difference of x-ray reflectivity when reversing chirality of circularly polarized x-rays (a) and the direction of the magnetization (b).*

Magnetization-induced change in x-ray reflectivity can be obtained by comparing the reflectivity spectra recorded with the x-ray beam having opposite circular polarization or the spectra measured with a circularly polarized beam when reversing the sample magnetization. SFigure 2 shows a comparison of two approaches under ambient conditions. The spectra at SFig 2a were recorded at the Swiss Light Source synchrotron, at the SIM beamline using the ReSoXS instrument by changing the polarization of x-rays. The spectra at SFig 2b were recorded at European XFEL during the time-resolved experiments where the magnetic contrast was obtained by reversing the sign of the applied magnetic field. Both approaches give a very similar magnetization-induced reflectivity change.

3) Time resolved optical data

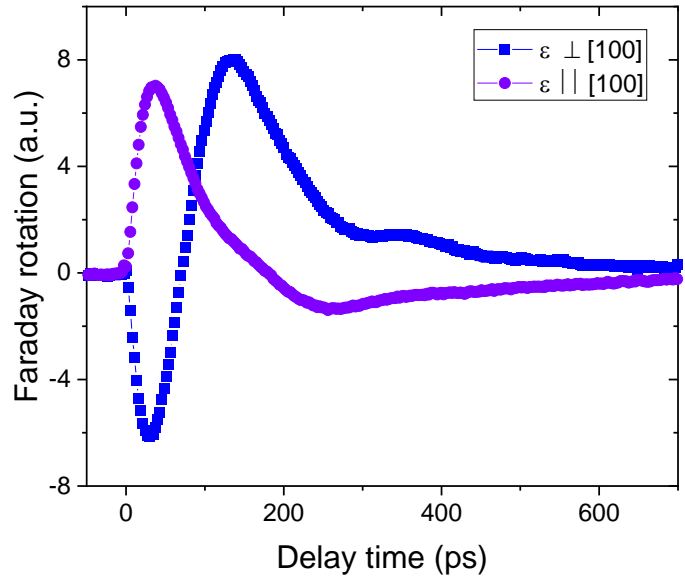

*SFigure 3. Time-resolved magnetization dynamics excited with 1300 nm pump wavelength and measured with 800 nm probe pulse in the same experimental conditions as during the x-ray probe experiments for two directions of pump polarization **ε** and external magnetic field H=250 Oe applied along [010] direction in YIG:Co film.*

SFigure 3 shows the transient Faraday rotation signal, reflecting the magnetization dynamics measured with the NIR pump-probe experiment in the same experimental configuration as the experiment with the x-ray probe but probed with 800 nm optical pulses. Optical excitation changes the preferred magnetization direction launching the magnetization dynamics. In the absence of an external magnetic field, the magnetization vector points along the easy axis defined by the laser excitation after ~ 30 ps. Making the optical excitation in an external magnetic field induces the same photoinduced change in the material that induces the magnetization switching. However, a stable switched state is not achievable due to the action of the external magnetic field that drives the magnetization dynamics after ~30 ps delay time.

4) Shot by shot fluctuation considerations

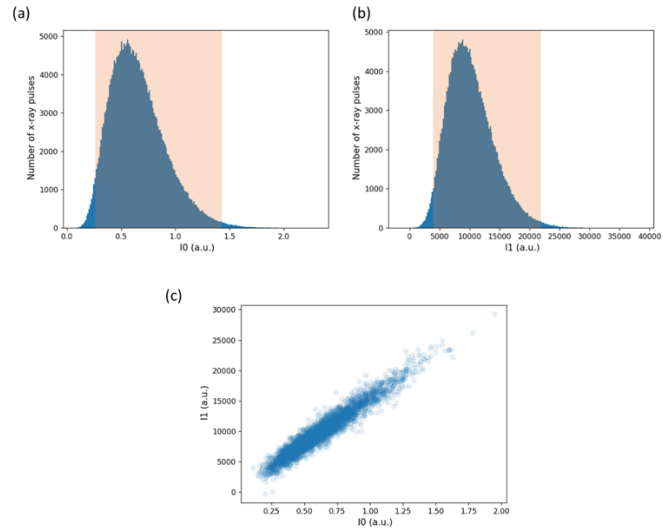

*SFigure 4. Example histogram of incoming x-ray pulse intensity I0 (a) and reflectivity from the sample I1 (b). The shaded area shows the interval that was used for the extraction of the time traces. Panel (c) shows the correlation between I1 and I0.*

The incoming pulse-to-pulse x-ray intensity fluctuation is very high after the monochromator. SFigure 3a and 3b show a histogram of incoming pulse x-ray intensities under equivalent conditions and the corresponding histogram for the x-ray reflected intensities, respectively. The data corresponds to a 10 min acquisition time or ~210000 x-ray shots. Despite quite significant shot-to-shot fluctuations of the intensity, we obtained a reasonable correlation between the two signals, as shown at SFig 3c.

## 5) Data reduction

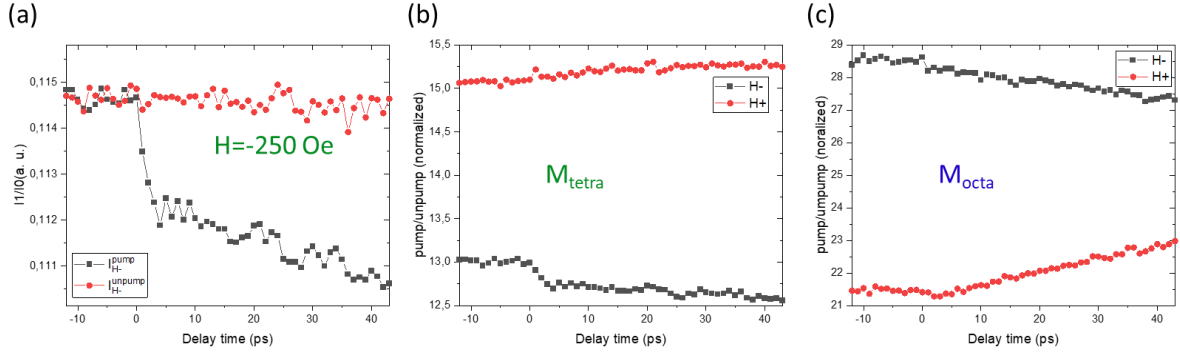

*SFigure 5. (a) Pumped and unpumped time-resolved reflectivity signal for H=-250 Oe magnetic field, recorded at a photon energy E=708.5 eV, which is sensitive to sublattice magnetization formed by the tetrahedrally coordinated Fe ions. (b) and (c) Time-dependent reflectivity for opposite directions of the external magnetic field probed with photon energies corresponding to the tetrahedral and octahedral magnetic sublattices, respectively. The time-resolved signals are normalized to unpumped values and vertical offset is applied to match the static x-ray reflectivity at negative delay time for the respective photon energies and direction of the external magnetic field (see SFig 1b) to show the evolution of x-ray reflectivity difference over the time after the excitation. The pump polarization was parallel to the [100] direction for all panels corresponding to the configuration, presented in Fig. 2c and 2d.*

The repetition rate of the x-ray pulses was 113 kHz, which is twice the optical pulses with 56.5 kHz. The normalization to unpumped signals allows us to suppress possible long-time drifts. An example of the time evolution of pumped and unpumped signals are shown in SFig 4a. With such a normalization, we track the evolution of the signal with respect to its unperturbed value. The change of a magnetization state is proportional to the difference between two signals at the opposite magnetic field. After normalization to unpumped, the value of the signal should be one for negative delay times. The difference between the two dependences recorded with the opposite magnetic field is that the signal for tetrahedral sublattice increases with time (SFig 4b) and decreases for octahedral (SFig 4b), meaning that projection of $M_{tetra}$ moment onto x-ray direction increases for $M_{octa}$ is decreasing.

## 6. Model of noncollinearity and spin dynamics in YIG:Co with low net magnetization

In order to explain the spin dynamics of a two-sublattice ferrimagnet in the vicinity of a magnetic compensation point (quasi-antiferromagnetic state), we use the Lagrangian $L_{ef}$ and Rayleigh dissipation functions $R_{ef}$ given in Ref. [1]. In this case, it is conventional to introduce the Néel vector $\vec{l} = \frac{\vec{M_1}-\vec{M_2}}{2M_O}$, an average sublattice magnetization $M_O = \frac{M_1+M_2}{2}$ and the dimensionless magnetization vector $\vec{m} = \frac{\vec{M_1}-\vec{M_2}}{M_O}$, which characterizes the proximity to the compensation point [1,2]. The magnetization of the tetrahedral and octahedral sublattices in YIG:Co films are $M_1 = M_{tetra}$ and $M_2 = M_{octa}$, respectively. In order to determine the spin dynamics, the spherical system of coordinates was used. Here, $\vec{l} = (sin\theta cos\varphi, sin\theta sin\varphi, cos\theta)$ and $\vec{M_i} = M_i(sin\theta_i cos\varphi_i, sin\theta_i sin\varphi_i, cos\theta_i)$, where $i$=1 or 2. In the quasi-antiferromagnetic approximation [2-4], the canting angles $\varepsilon$ and $\beta$ for tetrahedral and octahedral sublattices with respect to net magnetization are assumed to be small $\ll 1$. They determine the degree of noncollinearity of the sublattice magnetizations and are relevant here.

$$L_{eff} = \frac{\chi_\perp}{2}\left(\left[\frac{\dot\theta}{\gamma} + Hsin\varphi\right]^2 + \left[\frac{\dot\varphi}{\gamma}sin\theta + Hcos\theta cos\varphi\right]^2\right) - \frac{\dot\varphi}{\gamma}M_s cos\theta - U_A(\theta,\varphi) + M_s Hsin\theta sin\varphi - U_{eff}(\theta,\varphi), \quad (1)$$

$$R_{ef} = \frac{\alpha M_O}{2\gamma}(\dot\theta^2 + sin^2\theta\dot\varphi^2),$$

$$\theta_1 = \theta - \varepsilon, \theta_2 = \pi - \theta - \varepsilon, \varphi_1 = \varphi + \beta, \varphi_2 = \pi + \varphi - \beta, \chi_\perp = \frac{M_O}{H_e},$$

where $\chi_\perp$ is the transverse susceptibility of the garnet in the vicinity of compensation point [5], $U_A$ and $U_{eff}$ are the energies of the magnetocrystalline and photo-induced anisotropy, respectively [6], $\gamma$ is the gyromagnetic ratio, $\alpha$ is the Gilbert damping, $H_e$ is the exchange field, $M_s = mM_O = 4.5$ Gs is the magnetization saturation in the garnet, $\theta$ and $\varphi$ are the coordinates angles in the polar coordinate system, and $H$ is the external in-plane magnetic field defined in Fig.1. The low magnetization saturation depends on the Ge concentration in YIG:Co films [7]. For one Ge per formula unit the exchange stiffness is significantly larger compared to pure YIG and the magnetic ordering is close to the compensation of the magnetic moment [8]. Thus we could expect that the quasi-antiferromagnetic state describes the situation well for YIG:Co. The equilibrium state of this system can be obtained by a minimization of the total energy. The amplitude of the noncollinearity will be significant for magnetization orientation between the ground state and the collinear state defined by the external magnetic field with $0 < H < 1.5$ kOe. We note that in our experiment we used an external magnetic field of 250 Oe. The spin-dynamics excited by the laser pulse can be described by the angles $\theta(t) = \theta_O + \theta_1(t)$ and $\varphi(t) = \varphi_0 + \varphi_1(t)$, where $\theta_1 \ll 1$, $\varphi_1 \ll 1$.

The resulting Euler-Lagrange equations describing the spin-dynamics have the form:

$$\ddot{\theta} + \alpha\omega_e\dot{\theta}_1 + \omega_1^2\theta_1 - \omega_e \sin\theta_O \dot{\varphi}_1 = S_1(t),$$
$$\ddot{\varphi}_1 + \alpha\omega_e\dot{\varphi}_1 + \omega_2^2\varphi_1 + \omega_e \frac{\dot{\theta}_1}{\sin\theta_O} = S_2(t), \quad (2)$$
$$\omega_1^2 = \frac{3}{2}\omega_a\omega_e \sin^2\theta_O \cos^2\theta_O,$$
$$\omega_2^2 = \frac{1}{2}\omega_a\omega_e \sin^2\theta_O,$$

with $\omega_e = \gamma H_e = 3.41$ THz is exchange frequency [8], $\omega_a = \gamma 2K_1/M_S$ is frequency of FMR mode ($K_1$ is the cubic anisotropy constant [6]), $S_1$ and $S_2$ are the spin-torques for two sublattices. The frequency spectrum of the spin dynamics must satisfy the characteristic equation:

$$(\omega_1^2 - \omega^2)(\omega_2^2 - \omega^2) = \omega^2 \omega_e^2 \quad (3)$$

Finally, from the biquadratic equation, the ferromagnetic ($\omega_{fmr}$) and quasi-antiferromagnetic ($\omega_{q-AFMR}$) mode frequencies can be obtained:

$$\omega_{fmr}^2 = \frac{\omega_1^2 + \omega_2^2 + \omega_e^2}{2} - \left(\left(\frac{\omega_1^2 + \omega_2^2 + \omega_e^2}{2}\right)^2 - \omega_1^2\omega_2^2\right)^{\frac{1}{2}}, \quad (4)$$

$$\omega_{q-AFRM}^2 = \frac{\omega_1^2 + \omega_2^2 + \omega_e^2}{2} + \left(\left(\frac{\omega_1^2 + \omega_2^2 + \omega_e^2}{2}\right)^2 - \omega_1^2\omega_2^2\right)^{\frac{1}{2}}. \quad (5)$$

1. A. Zvezdin, A. Kimel, D. Plokhov, and K. Zvezdin, *J. Exp. Theor. Phys.* 131, 130 (2020).
2. M.D. Davydova, K.A. Zvezdin, J. Becker, A.V. Kimel, A.K. Zvezdin, *Phys.Rev.B* 100, 064409 (2019).
3. T.G.H. Blank , K.A. Grishunin , E.A. Mashkovich , M.V. Logunov, A.K. Zvezdin, A.V. Kimel, *Phys.Rev.Lett.* 127, 037203 (2021).
4. A.K Zvezdin, M.D. Davydova, K.A. Zvezdin, *Physics Uspekhi* 61 (11) 1127 (2018).
5. Landolt-Börnstein, Numerical Data and Functional Relationships in Science and Technology, New Series, Group III, vol 12, Berlin: Springer-Verlag, (1978).
6. A. Stupakiewicz, K. Szerenos, M. D. Davydova, K. A. Zvezdin, A. K. Zvezdin, A. Kirilyuk and A. V. Kimel, *Nat. Comm.,* 10, 612 (2019).
7. J. Šimšová, I. Tomaš, P. Görnert, M. Nevřiva, and M. Maryško, phys. stat. sol (a), 53, 297 (1979).
8. A. Gerhardstein et al *Phys.Rev.B* 18, 2218 (1978).